\documentclass[aps,prx,twocolumn,floatfix,longbibliography,nofootinbib,superscriptaddress]{revtex4-1}
\usepackage[autostyle]{csquotes}
\usepackage[utf8]{inputenc}
\usepackage{natbib}
\usepackage{graphicx}
\usepackage{xcolor}
\usepackage[tbtags]{amsmath}
\usepackage[colorlinks, linkcolor=blue]{hyperref}
\hypersetup{colorlinks,allcolors=black}
\usepackage{titlesec}
\usepackage{amssymb}
\usepackage{gensymb}
\usepackage{float}
\usepackage{braket}
\usepackage{tabularx,graphicx}
\usepackage{epstopdf}
\usepackage{latexsym}
\usepackage{color, colortbl}
\usepackage{psfrag}
\usepackage{txfonts}
\usepackage{bbm}
\usepackage{bm}
\usepackage{dsfont}
\usepackage{feynmp}
\usepackage{slashed}
\usepackage{multirow}
\usepackage[normalem]{ulem}

\makeatletter
\titleformat{\section}[block]{\normalfont\bfseries\centering}{\thesection}{1em}{\MakeUppercase}

\renewcommand{\appendix}{%
  \par
  \setcounter{section}{0}%
  \renewcommand{\thesection}{\Alph{section}}
  \renewcommand{\theHsection}{appendix.\Alph{section}}
  \titleformat{\section}[block]{\normalfont\bfseries\centering}{Appendix \thesection:}{1em}{}%
}
\makeatother

\renewcommand{\vec}[1]{\boldsymbol{#1}}

\def \R {{\cal{R}}}

\def \beq {\begin{eqnarray}}
\def \eeq {\end{eqnarray}}
\def \tn {\textnormal}

\definecolor{applegreen}{rgb}{0.5, 0.8, 0}

\begin{document}
\title{Quasicrystalline Spin Liquid}
\author{Sunghoon Kim}
\affiliation{Department of Physics, Cornell University, Ithaca NY 14853, U.S.A.}
\author{Mohammad Saad}
\affiliation{Department of Physics, Indian Institute of Technology Kanpur, Uttar Pradesh 208016, India.}
\author{Dan Mao}
\affiliation{Department of Physics, Cornell University, Ithaca NY 14853, U.S.A.}
\author{Adhip Agarwala}
\affiliation{Department of Physics, Indian Institute of Technology Kanpur, Uttar Pradesh 208016, India.}
\author{Debanjan Chowdhury}
\affiliation{Department of Physics, Cornell University, Ithaca NY 14853, U.S.A.}
\begin{abstract}
The interplay of electronic interactions and frustration in crystalline systems leads to a panoply of correlated phases, including exotic Mott insulators with non-trivial patterns of entanglement. Disorder introduces additional quantum interference effects that can drive localization phenomena. Quasicrystals, which are neither disordered nor perfectly crystalline, are interesting playgrounds for studying the effects of interaction, frustration, and quantum interference. Here we consider a solvable example of a quantum spin liquid on a tri-coordinated quasicrystal. We extend Kitaev's original construction for the spin model to our quasicrystalline setting and perform a large scale flux-sampling to find the ground-state configuration in terms of the emergent majorana fermions and flux excitations. This reveals a fully gapped and time-reversal symmetric quantum spin liquid, regardless of the exchange anisotropies, accompanied by a tendency towards non-trivial (de-)localization at the edge and the bulk. The advent of moir\'e materials and a variety of quantum simulators provide a new platform to bring phases of quasicrystalline quantum matter to life in a controlled fashion.
\end{abstract}
\maketitle

\section{Introduction}
Quasicrystals represent a fascinating and unique form of atomic arrangement \cite{ranganathan1991quasicrystals,Levine_PRL_1984, Goldman_RMP_1993, goldman1991quasicrystal}, where the sites are neither perfectly periodic, as in a regular crystal, nor maximally disordered, as in an amorphous material.  They can display Bragg-like peaks, but with ``forbidden'' (e.g. five-fold) symmetries, and provide an interesting playground to study the effect of strong local-interactions. Since their original discovery in Al-Mn alloys \cite{Levine_PRL_1984}, and subsequently in naturally occurring minerals \cite{stein}, recent years have brought the mysteries of this subject to the forefront with new experiments on highly tunable two-dimensional quasicrystals in the moir\'e setting \cite{uri2023superconductivity, ahn2018dirac, pezzini202030, deng2020interlayer, lv2021realization}. On the theoretical front, examples of novel quasicrystalline phases without any crystalline analogs have been discussed \cite{Varjas_PRL_2019,Else_PRX_2021,fan2022topological,Longhi_PRL_2019,cain2020layer,Jeon_PRB_LL,DuncanPRB2020}, especially in the non-interacting limit where the appearance of robust zero modes has been highlighted  \cite{Kohmoto_PRB_1987, Kohmoto_PRB_1986, Kohmoto_PRL_1986, Arai_PRB_1988, kohmoto1987electronic}. The properties of correlated electrons and of frustrated local-moment systems, in a quasicrystalline environment has not received as much attention. A series of recent works \cite{Kim_amorph_prl,Grushin_amorph_prl,Cassella_2023} have explored the effects of strong interactions in two-dimensional amorphous systems. There is a long history of studying frustrated spin models on a variety of quasicrystals; see e.g. \cite{QCfrust1,QCfrust2,AJ1,AJ2,QCfrust3} and references therein.

\begin{figure}[h!]
\centering
\includegraphics[width=0.45\linewidth]{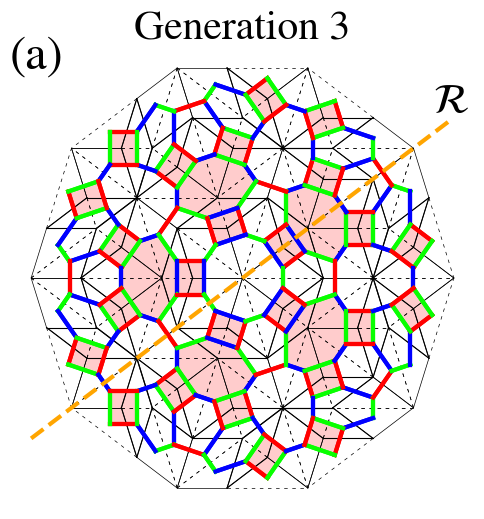}
\includegraphics[width=0.5\linewidth]{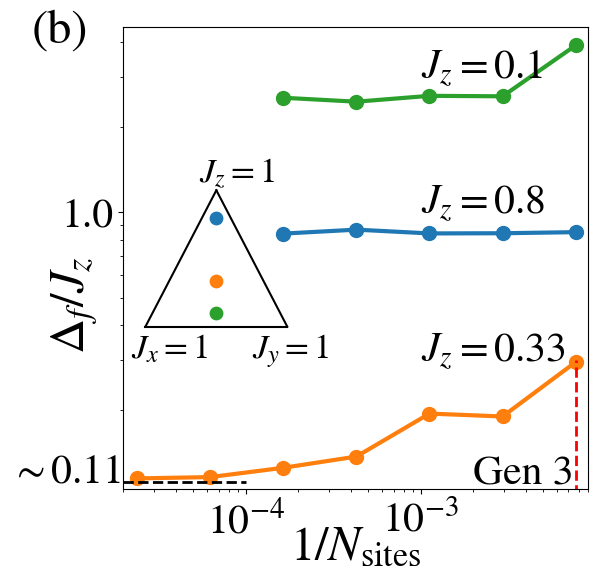}
\includegraphics[width=0.45\linewidth]{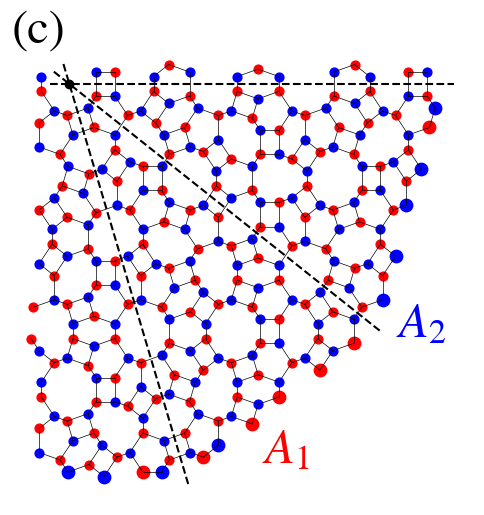}
\includegraphics[width=0.47\linewidth]{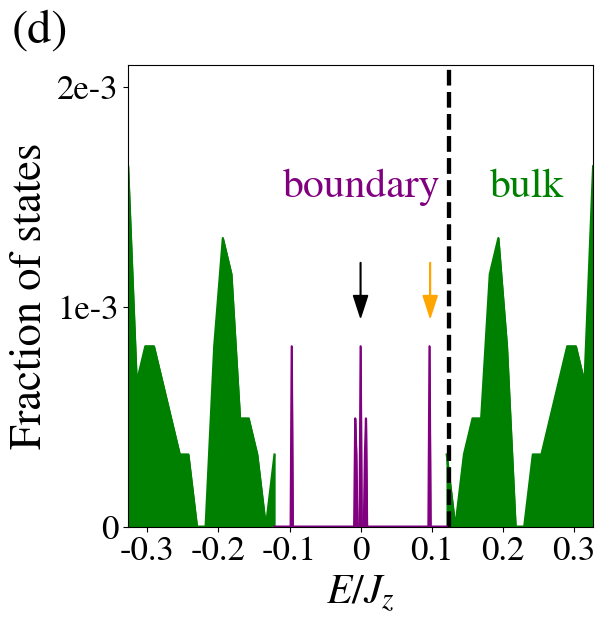}
\caption{(a) Tri-coordinated quasicrystal with a specific RBG-coloring after generation $\#3$. The original Penrose quasicrystal (black solid lines) divided into two identical triangles using the black dashed lines. The ground state in the canonical flux sector hosts $\pi$ flux on the squares and octagons (shaded plaquettes). The orange dashed line marks one of the five reflection planes, $\R$. (b) Finite size scaling of the bulk gap, $\Delta_f/J_z$, for three representative points in the parameter space; see inset. The smallest system-size corresponds to generation $\#3$; the extrapolated value ($N_{\rm{sites}}\rightarrow\infty$) of the bulk gap at the isotropic point is $\Delta_f/J_z\sim0.11$. (c) Sublattice sites, $A_1$ and $A_2$, with enlarged boundary dots highlighting the local imbalance (see text). Reflection planes are shown as dashed lines. (d) Low-energy density of states for generation $\#7$ with the lowest-$E$ bulk state (black dashed line) and two representative boundary states (arrows).}
\label{fig:qc_config}
\end{figure}

In this work, we construct an exactly solvable model of a quantum spin liquid on a {\it tri-coordinated} quasicrystal. Specifically, our model is a generalization of the celebrated Kitaev-model \cite{Kitaev_2006} for spin-$\frac{1}{2}$ degrees of freedom on the quasicrystal, instead of the usual honeycomb lattice; see Refs.~\cite{ARCMP,trebst} for possible material explorations of Kitaev spin liquids. As we discuss in detail below, many aspects of our ground-state phase diagram will be distinct from the usual Kitaev spin liquid (and its various disordered versions, which have been studied extensively \cite{dis1,dis2,dis3,dis4}) as a result of the underlying quasicrystalline character. While the low energy emergent excitations remain Majorana fermions and $Z_2$ fluxes, their spectra --- in the bulk and at the edge --- are distinct from the standard results for their crystalline counterparts. All of our results rely on an extension of Lieb's theorem \cite{LiebPRL1994,Macris_1996} in the unusual quasicrystalline setting, and helps simplify the numerical diagonalization of the interacting Hamiltonian. Notably, our quasicrystalline spin liquid ground state is {\it not} flux free and (on average) contains an irrational flux per unit area. Moreover, the Majorana fermions are non-uniformly localized near the edge, exhibiting a five-fold rotational symmetry, while the bulk excitations remain gapped regardless of the choice of bond couplings. Our exactly solvable spin liquid serves as a useful and possibly first example of a time-reversal symmetric quasicrystalline spin liquid for spin-$\frac{1}{2}$ degrees of freedom, with a non-trivial pattern of (de-)localization along the bulk and edge as a function of energy.

\begin{figure*}
\includegraphics[width=0.43\linewidth]{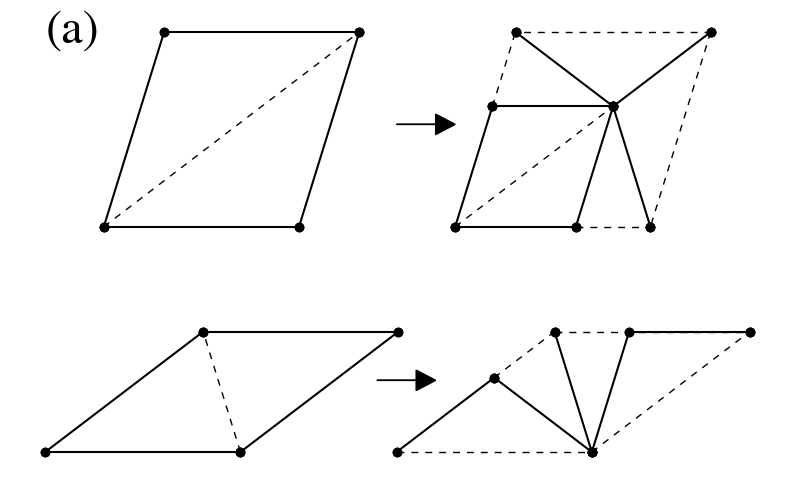}
\includegraphics[width=0.28\linewidth]{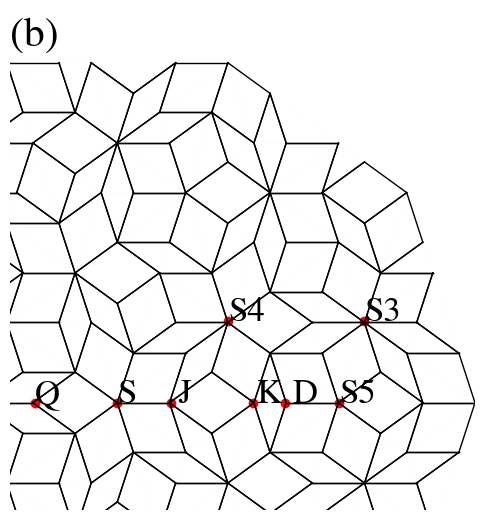}
\includegraphics[width=0.28\linewidth]{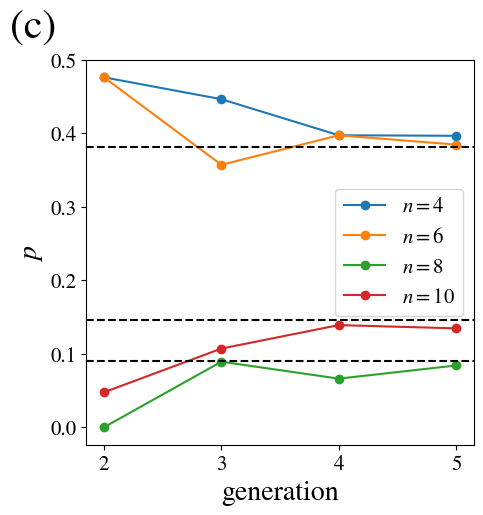}
\caption{(a) Illustration of inflation rules for the Penrose QC. Upon inflation, each rhombus is divided into smaller rhombi. The solid lines mark the boundary of rhombi, while the dashed lines serve as a guide to eye for triangulation. (b) Illustration of the 8 types of vertices of the Penrose QC. (c) The fraction of each polygon $(n=4, 6, 8, 10)$ with increasing generation. The black dashed lines mark the asymptotic expected behavior of $\tau^{-2}, \tau^{-4}, \tau^{-5}$ ($\tau\equiv$golden-ratio), respectively, from top to bottom. }
\label{fig:inflation}
\end{figure*}

\section{Quasicrystalline graph} Our starting point involves designing a tri-coordinated quasicrystal, which will allow us to label the bonds using a compass-like Ising interaction. We start with the celebrated ``Penrose tiling" consisting of two types of rhombii, each of which is made of two identical isosceles triangles --- the golden triangle and golden gnomon \cite{penrose1974role}. We obtain the tri-coordinated quasicrystal by connecting the centroids of the adjacent triangles; see Appendix~\ref{appdx:add_const}. The Penrose tiling is itself constructed in the usual way by following the standard inflation rules for successive generations \cite{Bruijn1981,Kumar_PRB1986}. Associated with every such generation, we obtain the corresponding tri-coordinated quasicrystal; see Fig.~\ref{fig:qc_config}(a) and Appendix~\ref{appdx:add_const} for a visualization of the  graph after generation $\#3$. The tri-coordinated quasicrystal is also bipartite, with a local ``imbalance" of the two sublattice sites at the edge (sites labeled as $A_1,~A_2$; see Fig.~\ref{fig:qc_config} (c)), and contains four types of polygons: square, hexagon, octagon, and decagon, respectively. The Kitaev model on this graph, as described in detail below, leads to a fully gapped spin liquid in the bulk (Fig.~\ref{fig:qc_config}(b)). 

Let us now expand further on the inflation rules. From the inflation properties of the Penrose QC (see Fig.~\ref{fig:inflation}(a)),
 we can obtain the fraction of each polygon in the thermodynamic limit. The Penrose QC consists of 8 types of vertices with different connectivities, which are denoted as $D, J, Q, K, S3, S4, S$ and $S5$; see Fig.~\ref{fig:inflation}(b). Let $\vec{n}_k$ denote an eight-dimensional vector for the number of each type of vertex for generation $\#k$. Upon inflation, two consecutive number vectors are related by a transfer matrix, i.e. $\vec{n}_{k+1}=M\vec{n}_k$ \cite{de1981algebraic,Kumar_PRB1986}. The fractions of each type of vertex in the thermodynamic limit are determined by the largest eigenvalue of the matrix, $\tau^2$, where $\tau=(\sqrt{5}+1)/2$ is the golden ratio. In the thermodynamic limit, the fraction of each type is given by $\left(\frac{1}{\tau^2},\frac{1}{\tau^3},\frac{1}{\tau^4},\frac{1}{\tau^5},\frac{1}{\tau^6},\frac{1}{\tau^7},\frac{1}{\tau^4 (1+\tau^2)},\frac{1}{\tau^6 (1+\tau^2)}\right)$. We can readily find the correspondence between the vertices and the polygons as follows: $D \leftrightarrow 4, (Q,J)\leftrightarrow6, K\leftrightarrow 8$, and $(S3,S4,S,S5)\leftrightarrow 10$. We thus obtain the fraction of each polygon in the thermodynamic limit as
\beq 
(p_4, p_6, p_8, p_{10})=\left(\frac{1}{\tau^2},\frac{1}{\tau^2},\frac{1}{\tau^5},\frac{1}{\tau^4}\right).
\eeq  
In Fig.~\ref{fig:inflation} (c), we present the fraction of each polygon as a function of generation. Each fraction converges to irrational values, $\tau^{-x}$ $(x=2,4,5)$, marked by black dashed lines.  

\section{Kitaev Model} We introduce a coloring scheme to label each of the three bonds around a vertex as red ($R$), blue ($B$), and green ($G$). Each such vertex is connected to its 3 neighboring sites via an $x$-, $y$-, and $z$-link (corresponding to R, B, G respectively) depending on the direction. These nearest-neighbor bonds represent Kitaev-type spin interactions for $S=1/2$ degrees of freedom on the above quasicrystalline graph. 
The Hamiltonian is defined as
\beq 
H=\sum_{\langle i,j \rangle_\mu}J^{\mu}_{ij}\sigma_{i}^{\mu}\sigma_{j}^{\mu},
\label{eq:Kitaev_spin}
\eeq 
where $\sigma_i^\mu$ $(\mu=x,y,z)$ is a Pauli operator for the spin at site $i=1,\dots ,2N$, $\langle i,j \rangle_\mu$ is a pair of nearest neighboring sites connected via a $\mu$-link, and $J^{\mu}_{ij}$ are link-dependent couplings. 
The coloring scheme is not unique, and we choose a specific convention whereby $\{R,B,G\}$ bonds are assigned in a clockwise (counter-clockwise) fashion for vertices belonging to the $A_1 \ (A_2)$ sublattice, respectively; see Fig.~\ref{fig:qc_config} (a) and  {Appendix~\ref{appdx:add_const}. Our coloring scheme manifestly preserves the {\it reflection symmetries} associated with the bond couplings, which will turn out to be crucial for application of Lieb's flux theorem \cite{LiebPRL1994,Macris_1996}. Interestingly, each polygon (plaquette) then consists of only two types of colored bonds, in contrast to the usual honeycomb model. \footnote{To avoid any spurious extrinsic edge effects due to the boundary sites with a coordination number of two, we introduce a scheme to pair them up by infinitesimal couplings and preserve their tri-coordinated nature; see Appendix~\ref{appdx:add_const}.}

We define the usual flux operators, $W_p \equiv \prod_{\langle i,j \rangle_\mu \in \partial P}\sigma_{i}^{\mu}\sigma_{j}^{\mu}$, where $\partial P$ denotes the boundary of the plaquette $p$ oriented in a clockwise fashion. Since $[H,W_p]=0$, the eigenstates of the Hamiltonian Eq.~\ref{eq:Kitaev_spin} can be divided into different flux sectors, $\{ w_p\}$, where $w_p =\pm 1$ represent the flux eigenvalues. Following Kitaev's presciption \cite{Kitaev_2006} we rewrite a spin operator in terms of Majorana fermions $\sigma^\mu_i =ib^\mu_i c_i$, and define link operators $\hat{u}_{ij} =ib_i^\mu b_j ^\mu$. Then Eq.~\ref{eq:Kitaev_spin} can be rewritten as $H_{\hat{u}}=i\sum_{\langle j,k \rangle_\mu}J^{\mu}_{jk} \hat{u}_{jk} c_j c_k$, 
with $[H_{\hat{u}},\hat{u}_{jk}]=0$, leading to the link operators having conserved eigenvalues. Since the link operators take eigenvalues $u_{jk}=\pm1$, the problem reduces to $c-$Majorana fermions coupled to the static $Z_2$ gauge fields $\{ u_{jk} \}$, hopping on the sites of a quasicrystal; however, the lowest energy many-body state need not correspond to all of the $\{ u_{jk} \}=1$.  For the translationally invariant honeycomb lattice, the ground state is known to be flux-free according to Lieb's flux theorem \cite{LiebPRL1994}. We next turn to obtaining the ground state flux configuration on the tri-coordinated quasicrystalline graph.

\section{Extended Lieb's theorem for ground state} Given that the quasicrystal consists of distinct polygons, it is not obvious {\it a priori} whether the ground state can be flux-free, and if not, which polygons host non-zero fluxes. To find the ground state configuration, we perform an extensive sampling of all the distinct flux configurations numerically. We leverage a ``generalized" version of Lieb's theorem \cite{Macris_1996, Jaffe_2014, chesi2013vortex} that reduces dramatically the number of samplings required to identify the ground state. 

To proceed, we first rewrite the Kitaev quadratic Hamiltonian as, 
\beq 
H_{u}=\frac{i}{2}(c_{A_1}^T \ c_{A_2}^T) 
\begin{pmatrix}
0 & M \\
-M^T & 0 
\end{pmatrix}
\begin{pmatrix}
c_{A_1} \\
c_{A_2} 
\end{pmatrix},
\label{eq:Kitaev_u}
\eeq 
where $M_{ij}=J^\mu_{ij}u_{ij}$, and $c_{A_1\:(A_2)}$ denotes the vector of Majorana operators for the sublattice $A_1\:(A_2)$. We perform a canonical transformation to rewrite Eq.~\ref{eq:Kitaev_u} as $H_u=\sum_{m=1}^{N} \varepsilon_m (2a^\dagger _m a_m -1)$ in terms of complex fermions, $a_m,~a_m^\dagger$, where $\varepsilon_m$ are non-negative eigenvalues of $M$. The cost of adopting the above representation is an enlarged Hilbert space; we project out the unphysical states based on the parity of the fermionic eigenmodes, $\prod_m (1-2n_m )$, where $n_m =a^\dagger_m a_m$ \cite{Yao_PRL_parity,Pedrocchi_PRB2011}; see Appendix~\ref{appdx:projection}. 

Consider a generic two-dimensional bipartite graph $\Lambda$, with a nearest-neighbor Hamiltonian defined for Majorana fermions, $H=\sum_{x,y\in \Lambda} i J_{xy} u_{xy} c_x c_y$, 
where $J_{xy}\geq 0$ and $u_{xy} = \pm 1$. The flux of a given plaquette $W_p$ can be written as $W_p = \prod_{(ij)\in \partial p } u_{ij}$, where $i$ ($j$) belongs to the $A_1$ ($A_2$) sublattices of the boundary of a plaquette $p$. At {\it half-filling} (of the complex fermion representation), flux sectors minimizing the ground state energy satisfy the following properties \cite{chesi2013vortex}: (i) the plaquettes intersecting $\R$ have a canonical flux $W_p = - (-1)^{L/2}$ where $L$ is the number of edges of the plaquette, and (ii) the remaining plaquettes have the same flux as their reflected plaquette, i.e. $W_p = W_{r(p)}$.

For our quasicrystal, there are five reflection planes ($\R$) dividing the system into two mirror-symmetric halves. They bisect the edges of the central decagon at right angles; see Fig.~\ref{fig:qc_config}(a). The flux associated with the plaquettes intersecting $\R$ are fixed by property (i). For the remaining plaquettes, we divide the entire system into ten reflection-symmetric sectors. Within each sector involving $N_p$ plaquettes not intersected by the $\R$, we can identify a total of $2^{N_p}$ distinct flux sectors. We find the configuration with the minimum ground state energy, $E_{\tn{GS}}$, after sampling over all distinct flux configurations, keeping reflection-symmetric $\{u_{ij}\}$ and constraining to the physical Hilbert space with $\sum_m n_m =0$. We observe that $E_{\tn{GS}}$ is minimized when \textit{every} plaquette has canonical flux (see definition above and Fig.~\ref{fig:qc_config} (a)); we dub this the canonical flux sector. By varying the anisotropy associated with the Kitaev exchange (i.e. ~$J^z/J, J^x=J^y=J$) we have confirmed that the canonical flux sector continues to remain the ground state; see Appendix~\ref{appdx:flux_sampling}. We have performed the numerical flux samplings up to generation $\#4$ (where the total number of sites$=340$) and confirmed that the above result continues to remain valid. For higher generations, the sampling procedure becomes increasingly computationally intense, and so we assume that the ground state will continue to conform to the canonical flux sector. We next turn to the nature of the excitations above the canonical ground state flux sector.

\section{Results}
\subsection{Fermionic excitations} The Hamiltonian defined on any finite system reveals an interesting excitation spectrum; see, for instance,  Fig.~\ref{fig:qc_config}(d) for the density of states (DOS) at the isotropic point. While a naive interpretation might lead us to conclude the existence of a gapless phase (with $E=0$) over a broad range of parameters, a careful analysis reveals that the bulk and boundary states need to be disentangled first in order to obtain the actual phase diagram.

\begin{figure}[]
\centering
\includegraphics[width=0.4423\linewidth]{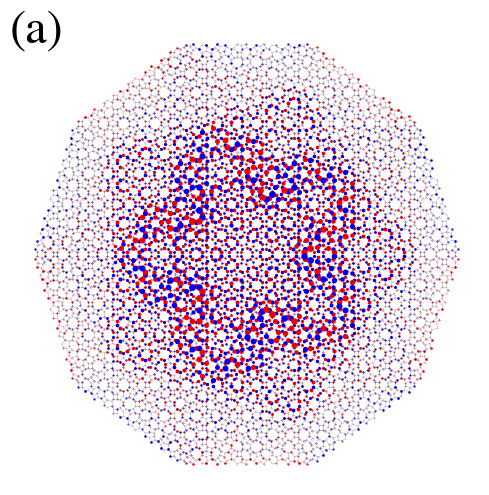}
\includegraphics[width=0.4423\linewidth]{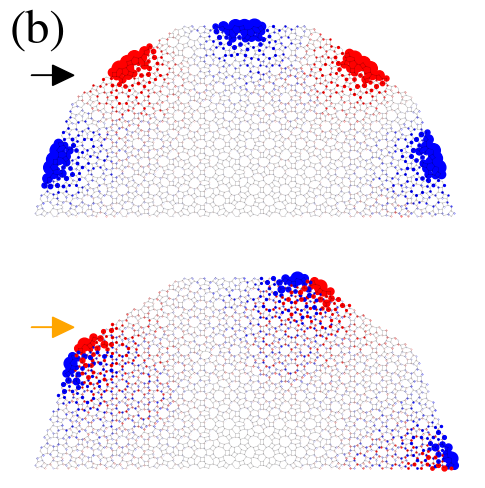}
\includegraphics[width=0.4422\linewidth]{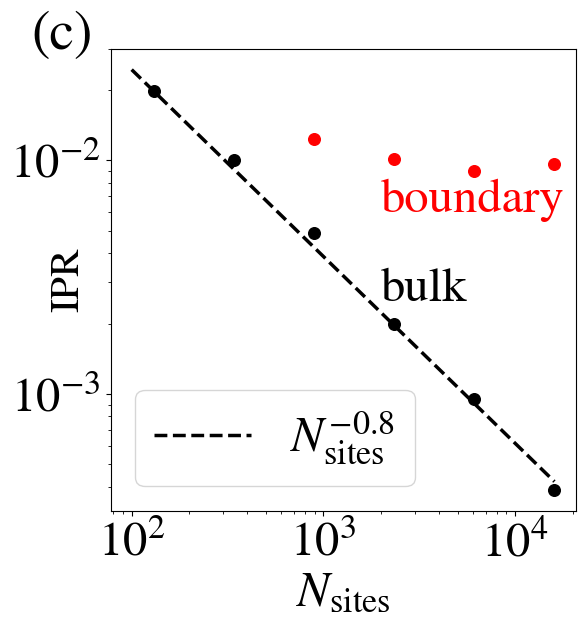}
\includegraphics[width=0.4422\linewidth]{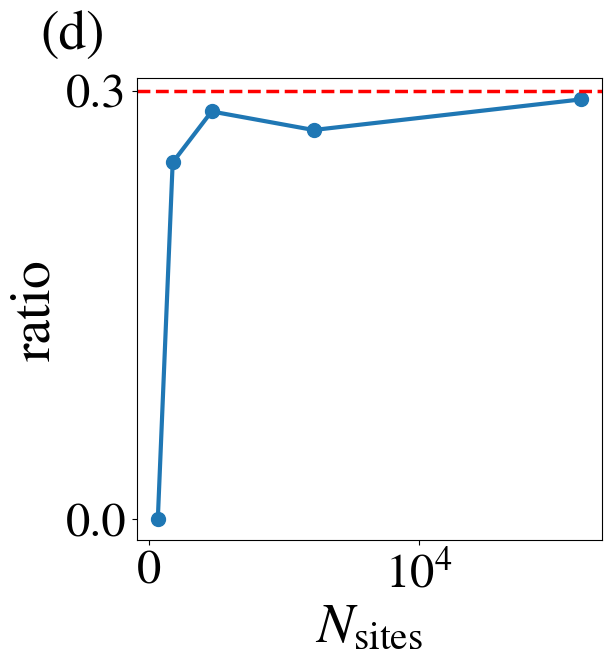}
\caption{Distribution of the fermionic wave function for (a) the lowest-$E$ bulk state, and (b) two representative boundary states associated with the marked arrows in Fig.~\ref{fig:qc_config}(d),  respectively. The blob size denotes the probability of finding the state at that site, and the color (red/blue) represents the sublattice. (c) Finite size scaling of the IPR. The bulk IPR is computed for the lowest-energy bulk state, while the boundary IPR is obtained by averaging over all boundary states. (d) The ratio between the number of boundary states and local imbalance, $\Delta N$, as a function of system size.}
\label{fig:boundary_bulk_wf}
\end{figure}

We first use the local DOS as a diagnostic. If the position of the peak of $|\psi_n(r)|^2$ is located deep in the bulk (near the boundary), where $\psi_n(r)$ is the wave function of the $n$-th state at site $r$, the state can be identified as a bulk (boundary) state; see Appendix~\ref{appdx:ldos} for details. We define the bulk gap $\Delta_f$ as the minimum energy eigenvalue of all the bulk states. Interestingly, we find that the fermionic spectrum is \textit{gapped} regardless of the choice of couplings, in stark contrast with the translationally invariant version of the original model on the honeycomb lattice. 

To demonstrate the gapped bulk spectrum, we perform finite-size scaling with a few representative points with different $J_z$,  keeping $J_x=J_y$ and $\sum_\mu J_\mu =1$ (Fig.~\ref{fig:qc_config}(b)). In one anisotropic limit $J_z \rightarrow 0$, our quasicrystal is decoupled into a set of disconnected polygons with only $x$ and $y-$links; $\Delta_f$ is thus determined by the spectrum of the largest polygon --- a decagon. Note that the same limit describes a set of decoupled one-dimensional chains for the honeycomb Kitaev model. In the other limit of $J_z \rightarrow 1$, the QC is divided into decoupled dimers connected by $z-$links, and $\Delta_f$ is determined by the spectrum of a single dimer; this is also true for the honeycomb Kitaev model. These energy scales in the two anisotropic limits remain finite with increasing number of generations, and the spectrum is gapped. At the isotropic point, we have carried out a study with increasing system size and find that $\Delta_f$ converges to a small but finite value $\Delta_f/J_z \sim 0.11$; see Fig.~\ref{fig:qc_config}(b). Note that the fermions in the canonical flux sector are subject to an effective inhomogeneous \textit{emergent} magnetic field through specific plaquettes. As this magnetic field is switched off, the system enters the flux-free sector, but with a {\it greater} energy, where we find the spectrum is gapless with $\Delta_f$ scaling as $\sim N_{\tn{sites}}^{-\alpha}$ with $\alpha\approx0.6$, where $N_{\tn{sites}}$ denotes the total number of sites; see Appendix~\ref{appdx:flux_sampling}. This suggests that the fermionic gap is likely tied to the \textit{Hofstadter butterfly} problem on the quasicrystal due to the inhomogeneous emergent $Z_2$ fluxes of the ground state \cite{vidal}; connecting this to the gap-labeling conjecture for quasicrystals \cite{Johnson_1982,Bellissard_2005,Benameur_2007} remains an interesting open problem. We have also studied the effect of a perturbative magnetic field along the $[1,1,1]$ direction. Similar to the honeycomb model, a chiral spin liquid (CSL) can be realized by a small magnetic field near the isotropic point. An important difference of the quasicrystalline model is that the field-induced transition is now accompanied by a bulk gap closing, as shown in Fig.~\ref{fig:csl_gap}.

\begin{figure}[h]
\centering
\includegraphics[width=0.8\linewidth]{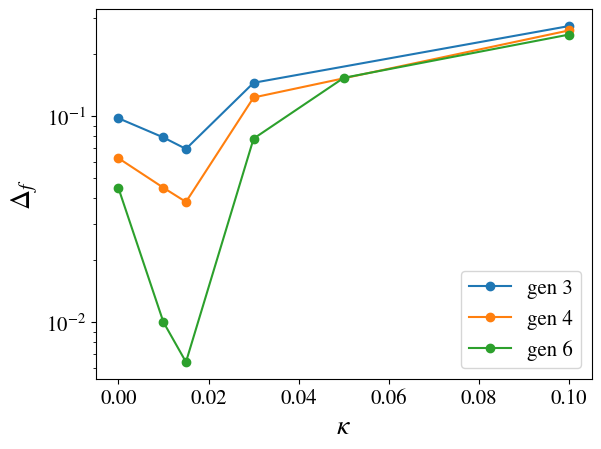}
\caption{Gap-closing transition from a gapped $Z_2$ spin liquid to a chiral spin liquid as a function of an external magnetic field $(\kappa\sim \frac{h_x h_y h_z}{J^2}\ll J)$ at the isotropic point.} 
\label{fig:csl_gap}
\end{figure}

Let us now comment on the spatial distribution of the bulk and boundary states tied to specific energies. We plot the wavefunction distribution for the lowest energy bulk state (Fig.~\ref{fig:qc_config}(d)) in Fig.~\ref{fig:boundary_bulk_wf}(a). It is evident that the wavefunction has support over a large fraction of the sites and exhibits a (quasi-)delocalized behavior, which will be clarified further when we introduce the inverse participation ratio (IPR) as a diagnostic. The boundary states associated with two different energies (Fig.~\ref{fig:qc_config}(d)) are clearly localized along the edge, and interestingly, have a strong sub-lattice dependence (Fig.~\ref{fig:boundary_bulk_wf}(b)). The boundary states are strongly affected by the local topology  \cite{Sutherland_PRB_localtopology, Day-Roberts_PRB2020} and favor the majority sublattice site along the boundary within each region (see Fig.~\ref{fig:qc_config} (c)). It is known that such \textit{local imbalance} between the population of two sublattices can result in localized zero-energy states \cite{Day-Roberts_PRB2020}. We revisit this point below, after highlighting the contrast between the (de-)localization properties associated with the bulk vs. boundary states.

\begin{figure*}
\includegraphics[width=0.23\linewidth]{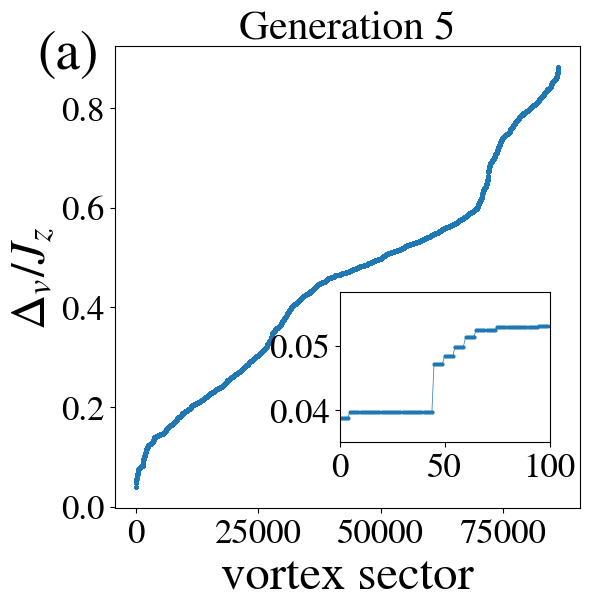}
\includegraphics[width=0.23\linewidth]{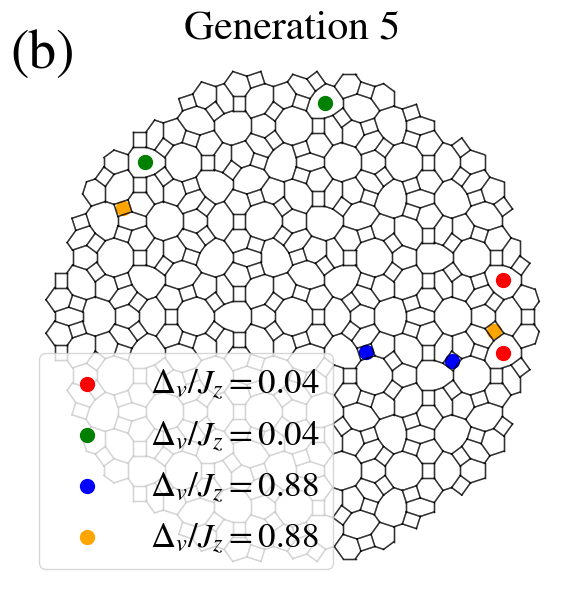}
\includegraphics[width=0.29\linewidth]{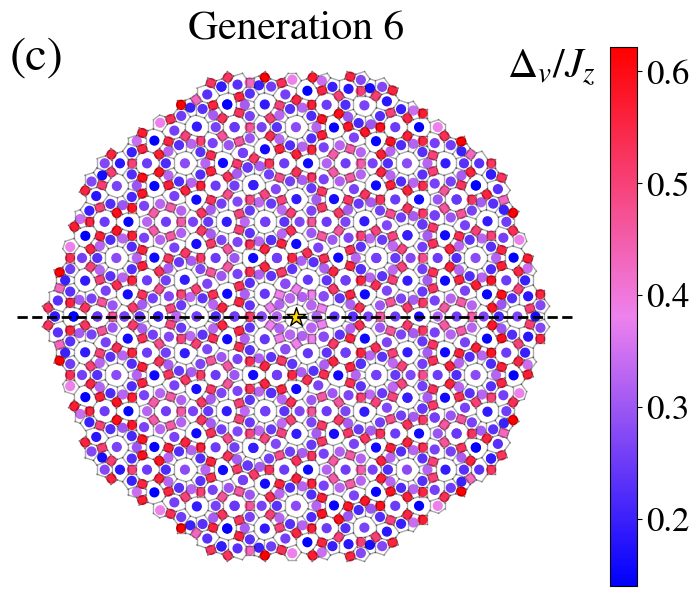}
\includegraphics[width=0.23\linewidth]{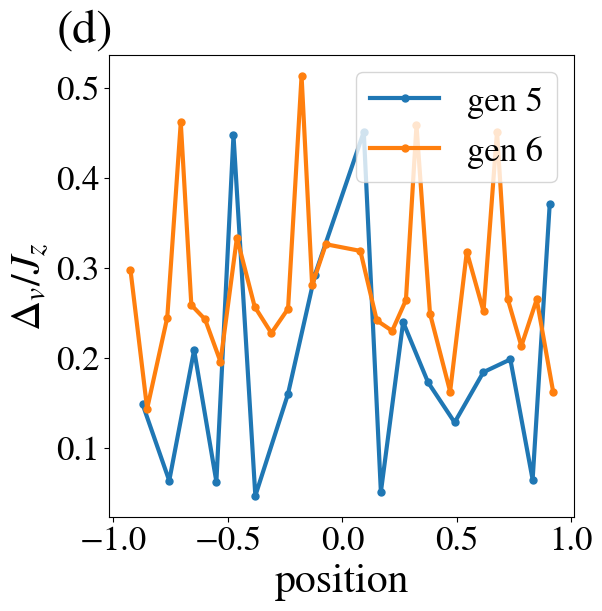}
\caption{(a) The vortex-pair excitation energy at the isotropic point for generation $\#5$. Inset: Enlarged plot for the first few plateaus. (b) The two lowest (red and green) and highest (blue and orange) energy vortex-pair configurations, respectively.  Note that vortices carry an $n\pi$ flux (mod $2\pi$) for the polygon with length $2n$. (c) The map of the vortex energy when one vortex is fixed at the center (denoted `$\star$'). (d) Position-dependence of the vortex energy along the horizontal dashed line shown in (c). The position is rescaled by the radius of the system.} 
\label{fig:vortex_pair_hierarchy}
\end{figure*}

Let us introduce the IPR as
\beq 
\tn{IPR}_n=\sum_{r=1}^{N_{\tn{sites}}} |\psi_n (r)|^4.
\eeq 
Recall that a typical extended state satisfies an IPR scaling as $\sim N_{\tn{sites}}^{-1}$, while a typical localized state has a system-size independent IPR. In Fig.~\ref{fig:boundary_bulk_wf}(c), we present the finite size scaling of the IPR for the lowest-energy bulk state and the IPR averaged over all the boundary states. As we noted in Fig.~\ref{fig:boundary_bulk_wf}(a), we indeed find that the bulk states are delocalized over a finite fraction of the system with an IPR scaling as $\sim N_{\tn{sites}}^{-0.8}$. The slight deviation from the typical extended behavior in crystalline systems with translational symmetry is controlled by the unconventional distribution of sites on the quasicrystal. The exponent is likely tied to the fractal dimension of the corresponding state \cite{Halsey_fractal,Kohmoto_critical,Roche_1997}.  
The boundary states, on the other hand, are strongly localized along the edges leading to a nearly size-independent IPR. To quantitatively study the relation between the boundary states and the local imbalance, we introduce $\Delta N=\sum_{\tn{regions}}|N_{A_1} -N_{A_2}|$, where $N_{A_1,A_2}$ represents the number of sites belonging to each sublattice in each of the 10 regions partitioned by the reflection planes (Fig.~\ref{fig:qc_config}(c)) \cite{Day-Roberts_PRB2020}. In Fig.~\ref{fig:boundary_bulk_wf}(d), we plot the ratio between the number of boundary states and $\Delta N$ with increasing system size. For generations $\#4$ $(N_\tn{sites}\leq 340)$ and higher, the ratio converges to a finite value $\sim 0.3$ (red dashed line), demonstrating the close connection between the boundary states and the local imbalance. Note that both the localized states and the local imbalance are restricted to the region near the boundary. The zero-energy states are fragile against perturbations that do not preserve the local imbalance. For the analysis of $E=0$ states, the $2-$coordinated sites are paired up by infinitesimal couplings to preserve the original boundary condition. Connecting these sites via a non-zero coupling eventually gets rid of these $E=0$ states. 

To illustrate the origin of the connection between local imbalance and zero-energy modes, consider a bipartite lattice with nearest-neighbor hoppings and a block-diagonal Hamiltonian, $H=
\begin{pmatrix}
0 & G \\
G^T & 0 
\end{pmatrix}$.
Due to the imbalance (e.g. without any loss of generality, more $A_1$ sites compared to $A_2$), the number of rows and columns in $G$ are different, leading to some eigenstates with zero eigenvalue. The corresponding $E=0$ states are then localized on the majority sublattice $(A_1)$. A prominent example of this phenomenon arises on the Penrose quasicrystal \cite{Kohmoto_PRL_1986,Arai_PRB_1988}, which hosts $E=0$ states throughout the bulk since the system can be partitioned into decoupled sectors by ``membranes" that preclude any zero-energy states \cite{Flicker_PRX_cdimer,Day-Roberts_PRB2020}. In contrast, for our tri-coordinated quasicrystal, only the boundary exhibits a local imbalance leading to localization along the edge. We have also obtained boundary states with a small non-zero energy, manifest in the intermediate peaks over a finite range of energy $0<E<\Delta_f$ (Fig.~\ref{fig:qc_config}(d)). These states remain distributed near the boundary, but are \textit{not} strictly localized.

\subsection{Vortex excitations} Turning our attention now to the vortex-like excitations, let us consider a pair of vortices on top of the canonical ground-state  flux sector at the isotropic point. In the absence of crystalline translational symmetry, the vortex pair energy can be strongly dependent on their locations. In Fig.~\ref{fig:vortex_pair_hierarchy}(a), we plot the complete set of vortex pair energy,  $\Delta_v$, for the vortices placed at any two polygons for generation $\#5$. The plot shows multiple plateaus (see inset), which represents the number of geometrically equivalent vortex configurations. In Fig.~\ref{fig:vortex_pair_hierarchy}(b), we display the vortex pair configurations associated with the two lowest and highest energies, respectively. Interestingly, the lowest energy vortex pair excitations correspond to the vortices located in a reflection-symmetric fashion, suggesting that a version of Lieb's flux theorem might still have some bearing on these low-energy excitations. Moreover, as already indicated above, in the absence of crystalline symmetries, $\Delta_v$ is not a simple monotonous function of the spatial separation between the vortices, but instead depends both on the type of polygon and their positions. To illustrate this, we fix one vortex at the center of the quasicrystalline graph and map out the $\Delta_v$ tied to placing the other vortex at any other location on the graph; see Fig.~\ref{fig:vortex_pair_hierarchy}(c). The energetic hierarchy indicates that $\Delta_v$ tends to be larger when a vortex is located on the smaller polygons. A scan of $\Delta_v$ along a high-symmetry cut (black dashed line in Fig.~\ref{fig:vortex_pair_hierarchy}(c)) is shown in Fig.~\ref{fig:vortex_pair_hierarchy}(d). The vortex pairs are \textit{deconfined} as the energy cost associated with moving them apart does not continue to grow with increasing separation.

\section{Outlook} This work demonstrates the complexity associated with the interplay of fractionalization, entanglement and (de-)localization in quasicrystalline graphs. While serving as an important proof-of-concept demonstration of the novel aspects of frustrated spin models in quasicrystals, the advent of programmable Rydberg quantum simulators and interesting dynamical protocols \cite{lukin} can potentially help realize such phases in the laboratory in the not too distant future. 

Construction of a lattice with a fixed coordination number has been recognized as a viable route toward a solvable spin liquid model in various non-crystalline systems \cite{Cassella_2023,Grushin_amorph_prl,Keskiner_qcqsl_prb}. However, unlike the example considered here, the above constructions inevitably host plaquettes with an odd number of edges. This leads generically to  chiral spin liquid phases associated with a spontaneously broken time-reversal symmetry, while the spin-liquid tied to the quasicrystal in this work includes only plaquettes with an even number of edges and no tendency towards spontaneous breaking of this symmetry. Moreover, our quasicrystalline graph possesses a forbidden rotational symmetry, not appreciated in previous constructions, which leads to an interesting interplay of localization-delocalization behavior as a function of energy on the low-energy excitation spectrum, which has not been appreciated in some of these previous works.

Going beyond the classic Kitaev model construction, other exciting avenues for future work include extending the model \cite{YK07,Yao_Lee_PRL} on the same (or different \cite{Keskiner_qcqsl_prb,arovas}) quasicrystals, studying variants of the Kitaev-Kondo model \cite{Tsvelik} inspired by intriguing experiments in metallic quasicrystals \cite{HFQC}, and addressing the properties of a few doped particles on top of the spin liquid ground state \cite{halasz}. Developing a field-theoretic understanding of these phases while incorporating their quasicrystalline character from the outset also remains an interesting open problem. Finally, studying the explicit role of disorder in the quasicrystalline spin liquid on inducing additional localization tendencies for the low-energy excitations and on their spectral gaps is an exciting future direction  \cite{therm_met_Laumann,therm_met_Self,Cassella_2023}.

\acknowledgements We thank Daniel Arovas,  Junmo Jeon and SungBin Lee for insightful discussions, and Omri Lesser for a critical reading of an earlier version of this manuscript. SK, DM and DC are supported in part by a New Frontier Grant awarded by the College of Arts and Sciences at Cornell University and by a Sloan research fellowship from the Alfred P. Sloan foundation to DC. DM is also supported by a Bethe postdoctoral fellowship at Cornell University. DC acknowledges the support provided by the Aspen Center for Physics, which is supported by National Science Foundation grant PHY-1607611. AA acknowledges support from IITK Initiation Grant (IITK/PHY/2022010). AA acknowledges engaging discussions at the Indo-French meeting ``Novel Phases of Matter in Frustrated Magnets" at Bordeaux, France.

\appendix 

\section{Additional details on the construction of the tri-coordinated quasicrystal}
\label{appdx:add_const}
\begin{figure}[h]
\includegraphics[width=0.3\linewidth]{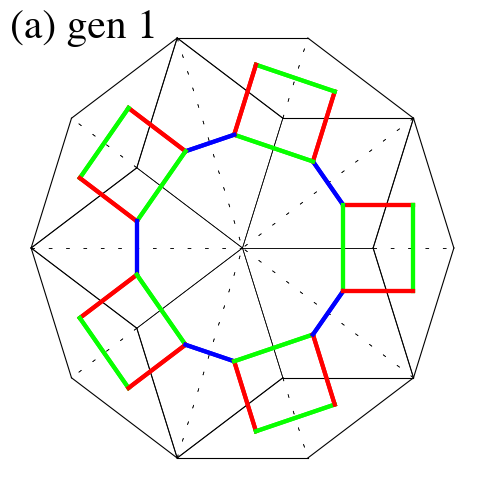}
\includegraphics[width=0.3\linewidth]{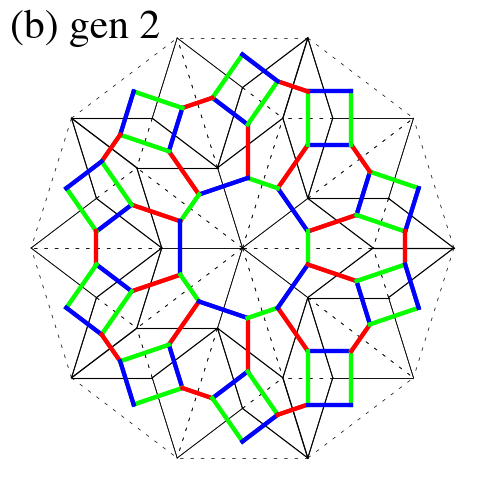}
\includegraphics[width=0.3\linewidth]{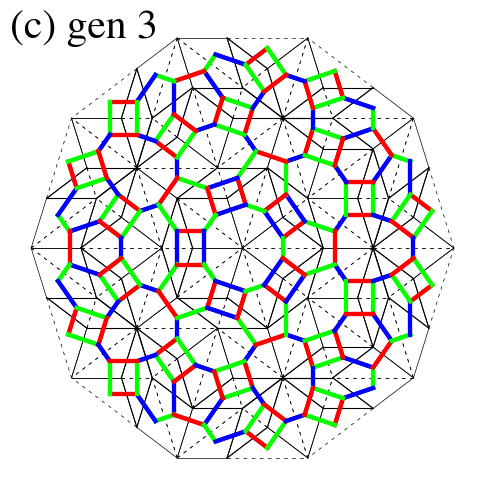}
\caption{Construction of our tri-coordinated quasicrystal starting from (a) generation $\#1$ up to (c) generation $\#3$. The underlying Penrose QCs are marked by black solid and dashed lines as with the main text. }
\label{fig:additional_construction}
\end{figure}

In Fig.~\ref{fig:additional_construction}(a)-(c), we present the configuration of the tri-coordinated quasicrystal up to generation $\#3$ with our specific choice of the 3-coloring convention (see main text) \cite{si}. The tri-coordinated quasicrystal can be viewed as the dual lattice of a triangulated Penrose quasicrystal of the same generation. We note an interesting even-odd effect tied to the configurations: The central decagon in the odd-generations is surrounded by 5 squares and 5 hexagons, while the central decagon in the even-generations is surrounded by 10 hexagons. This even-odd effect manifests in the finite size scaling of the bulk gap, e.g. see Fig.~\ref{fig:flux-free}.

\section{Projection onto the physical subspace}
\label{appdx:projection}
In this section, we provide details on  how to project out unphysical states, based on the derivation in \cite{Pedrocchi_PRB2011}. We consider a local operator, $D_i =-i\sigma_i^x \sigma_i^y \sigma_i^z=1$, which serves as the identity operator in the original physical subspace. We can rewrite the operator in terms of the Majorana operators as $D_i =b^x_i b^y_i b^z_i c_i$, which has eigenvalues $\pm1$. The projection operator onto the physical subspace is thus given by
\beq 
P=\prod_{i=1}^{2N}\left(\frac{1+D_i}{2}\right)=S\cdot \left(\frac{1+\prod_{i=1}^{2N}D_i}{2}\right) \equiv S\cdot  P_0,
\eeq 
where $S$ symmetrizes over all gauge equivalent $\{u_{ij} \}$  sectors and $P_0$ projects out unphysical states.

Our goal is to express $P_0$ in terms of the variables that are used to label a state: the field configuration $\{u_{ij}\}$ and the fermion occupation $\{n_m\}$. We start by rewriting $D\equiv \prod_{i=1}^{2N}D_i$ as
\beq 
D&=&\prod_{i=1}^{2N}c_i \prod_{i=1}^{2N}b_i^x \prod_{i=1}^{2N}b_i^y \prod_{i=1}^{2N}b_i^z \nonumber \\
&\equiv& (-1)^\theta \prod_{i=1}^{2N}c_i \cdot\prod_{\mu=x,y,z}\left( \prod_{\langle ij \rangle_\mu}b^\mu_i b^\mu_j \right) \nonumber \\
&=& (-1)^\theta \prod_{i=1}^{2N}c_i \cdot\prod_{\mu=x,y,z}\left( \prod_{\langle ij \rangle_\mu}-iu_{ij} \right),
\eeq 
where $(-1)^\theta=\pm1$ arises from the reordering of the $b_i^\mu$ operators, which depends on the choice of lattice labelings. For numerical calculations, we have chosen a labeling such that $i \leq N$ ($N+1\leq i$) belongs to the sublattice $A_1$ ($A_2$).

To deal with $\prod_{i=1}^{2N}c_i$, we revisit the canonical transformation of the Majorana Hamiltonian. We consider a singular-value decomposition, $M=USV^T$, where $U$ and $V$ are $N\times N$ orthogonal matrices and $S=\tn{diag}(\varepsilon_1,\dots,\varepsilon_N )$ with $\varepsilon_m \geq 0$. We rewrite the Hamiltonian as
\beq 
H_{u}&=&\frac{i}{2}(c_{A_1}^T \ c_{A_2}^T) 
\begin{pmatrix}
0 & M \\
-M^T & 0 
\end{pmatrix}
\begin{pmatrix}
c_{A_1} \\
c_{A_2} 
\end{pmatrix}\nonumber\\
&=&i\sum_{m=1}^{N}\varepsilon_m c_{A_1,m}'c_{A_2.m}' = \sum_{m=1}^{N}\varepsilon_m (2a^\dagger_m a_m -1),
\label{eq:canonical_ham}
\eeq 
where we have introduced new Majorana fermions, $(c_{A_1,1}',\dots,c_{A_1,N}')=(c_{A_1,1},\dots,c_{A_1,N})U$ and $(c_{A_2,1}',\dots,c_{A_2,N}')=(c_{A_2,1},\dots,c_{A_2,N})V$. The last equality in Eq.~\ref{eq:canonical_ham} is followed by the canonical transformation $a_m\equiv(c_{A_1,m}'+ic_{A_2,m}')/2$ and $a_m^\dagger\equiv(c_{A_1,m}'-ic_{A_2,m}')/2$. We then obtain 
\beq 
\prod_{i=1}^{2N}c_i &=& \prod_{m=1}^{N}c_{A_1,m} \prod_{m=1}^{N}c_{A_2,m} \nonumber\\
&=& \left(\tn{det}(U)\prod_{m=1}^{N}c_{A_1,m}'\right) \left(\tn{det}(V)\prod_{i=m}^{N}c_{A_2,m}'\right)\nonumber\\
&=&\tn{det}(U)\tn{det}(V)(-1)^{\frac{N(N-1)}{2}}\prod_{m=1}^{N}\left(c_{A_1,m}'c_{A_2,m}'\right).
\eeq 
Noting that $c_{A_1,m}'c_{A_2,m}'=i(1-2n_m)$, we finally obtain 
\beq 
D=(-1)^\theta \tn{det}(U)\tn{det}(V) (-1)^{\frac{N(N-1)}{2}} \cdot i^N \prod_{m=1}^{N}(1-2n_m) \cdot \left(\prod_{\langle ij \rangle_\mu} -i u_{ij} \right).
\label{eq:projection}
\eeq 
Eq.~\ref{eq:projection} can be used to determine $\{u_{ij}\}$ and $\{n_m\}$ for the physical ground state satisfying $P_0 =1$. For a given flux sector, a na\"ive ground state energy is $E_{gs}^{0}=-\sum_{m=1}^{N} \varepsilon_m$, which amounts to the energy of the \textit{half-filled} particle-hole symmetric spectrum, $(-\varepsilon_N,\dots,\varepsilon_N)$. If $\{u_{ij}\}$ from the canonical flux sector is physical when $\sum_{m=1}^{N}n_m =0$, then such a state is the true ground state according to the Lieb's theorem. If not, however, we have to change the parity of the fermion occupation by 1, departing from the half-filling. Thus, the flux sampling must extend to the ones violating the two basic assumptions (described in the main text) of the Lieb's theorem. For the flux sampling, one can come up with an appropriate spanning tree that bisects links of plaquettes (not intersected by the reflection planes) $N_p$ times. Each flux sector corresponds to a set of flipped links on the spanning tree. The resulting gauge choice for the canonical flux sector is, for instance, shown in Fig.~\ref{fig:spec_iso_gen4}.
\begin{figure}[h!]
\centering
\includegraphics[width=0.48\linewidth]{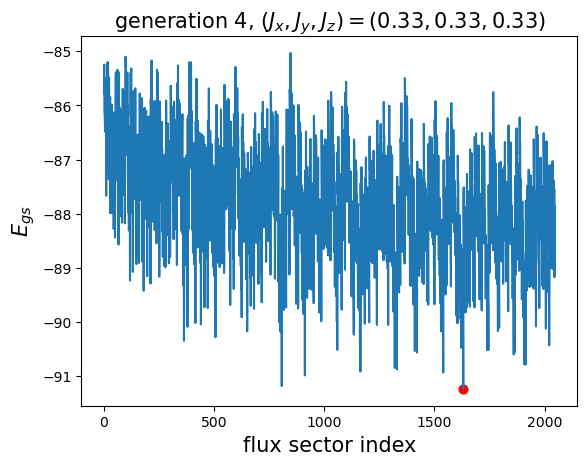}
\includegraphics[width=0.48\linewidth]{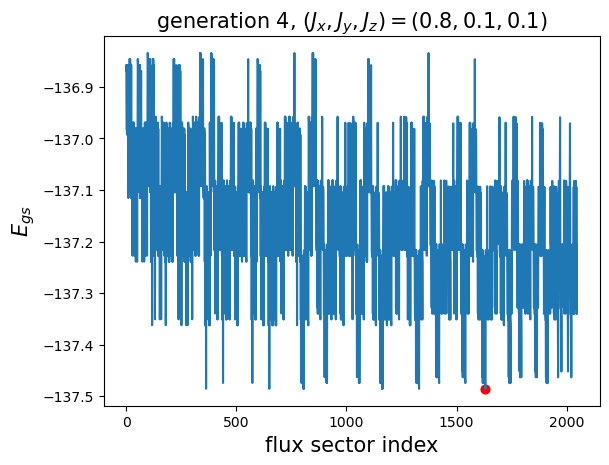}
\includegraphics[width=0.48\linewidth]{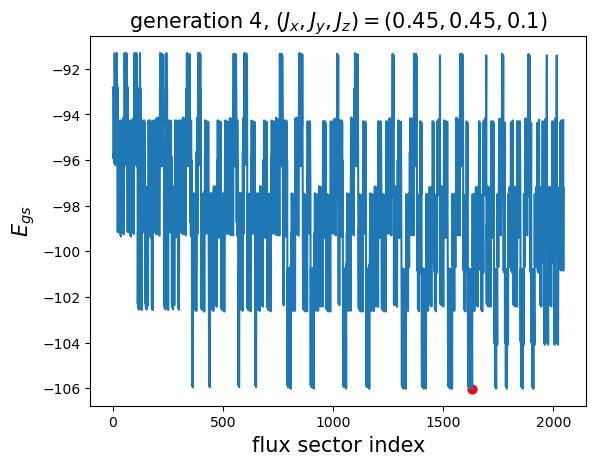}
\includegraphics[width=0.48\linewidth]{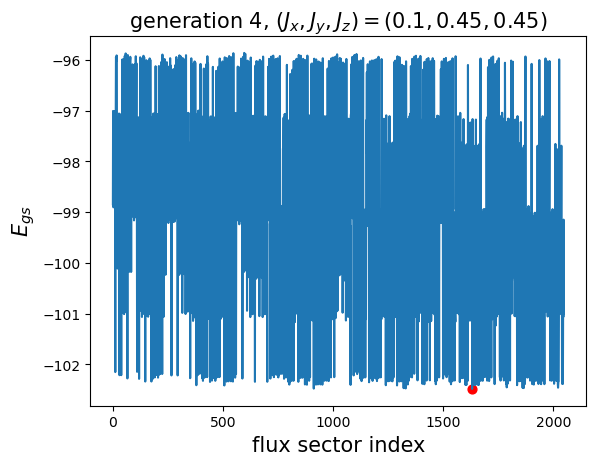}
\caption{Flux sampling for generation $\#4$ at four representative points in the phase diagram. The red-circle denotes the canonical flux-sector associated with the ground-state.}
\label{fig:samplings}
\end{figure}

\section{Additional results for flux sampling}
\label{appdx:flux_sampling}
\begin{figure}[h!]
\centering
\includegraphics[width=0.99\linewidth]{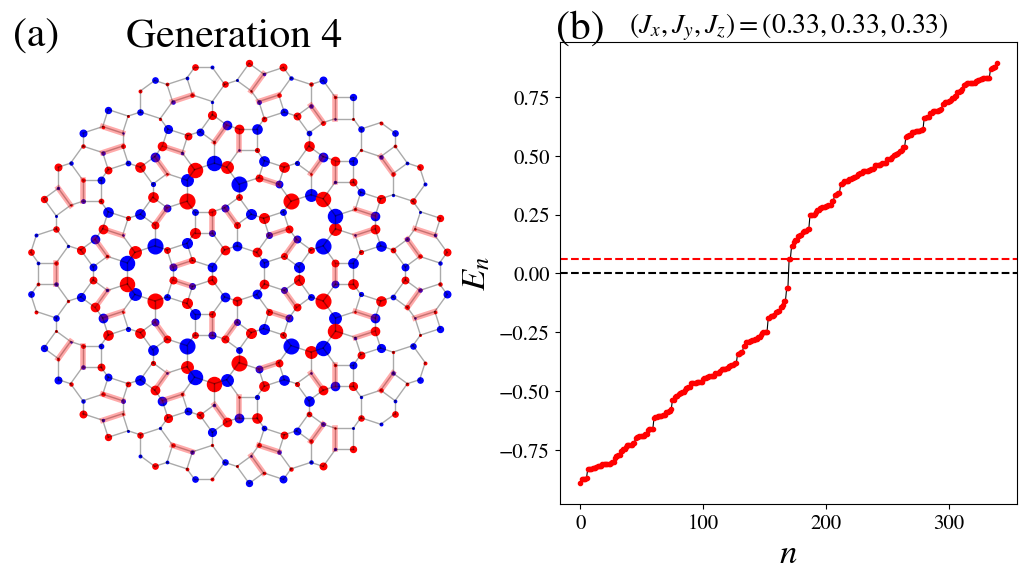}
\includegraphics[width=0.5\linewidth]{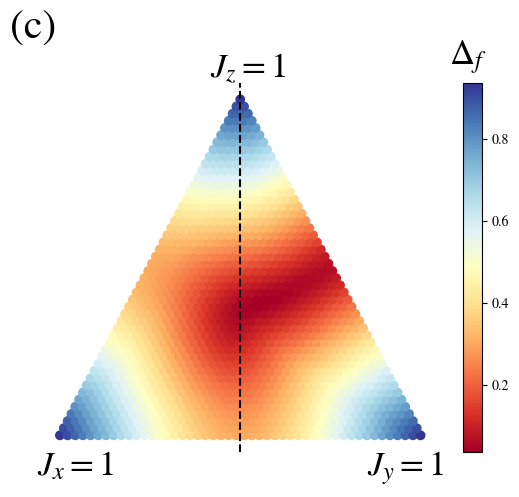}
\caption{(a) Ground state $Z_2$ field configuration and distribution of lowest-energy bulk fermionic excitations. The red (black) lines denote bonds with $u_{ij}=-1 \ (+1)$. The size of a blob quantifies the probability of finding the lowest energy fermionic excitations. (b) The fermionic spectrum with the red dashed line being the lowest-energy bulk state. (c) Bulk fermionic gap $\Delta_f$ on the plane $\sum_\mu J_\mu =1$ in the parameter space. The three representative set of parameters for the analysis of the bulk gap are located on the vertical dashed line.}
\label{fig:spec_iso_gen4}
\end{figure}

In Fig.~\ref{fig:samplings}, we present additional flux sampling results for generation $\#4$ at a few representative points in the phase diagram. The canonical flux sector (red circle) remains the ground state flux sector regardless of the choice of $\{J_\mu\}$. We show in Fig.~\ref{fig:spec_iso_gen4} the $Z_2$ field configuration for the ground state (canonical) flux sector at the isotropic point. Here the shaded red bond shows the links where $u_{ij}=-1$ in our gauge choice.

\begin{figure}[h]
\centering
\includegraphics[width=0.99\linewidth]{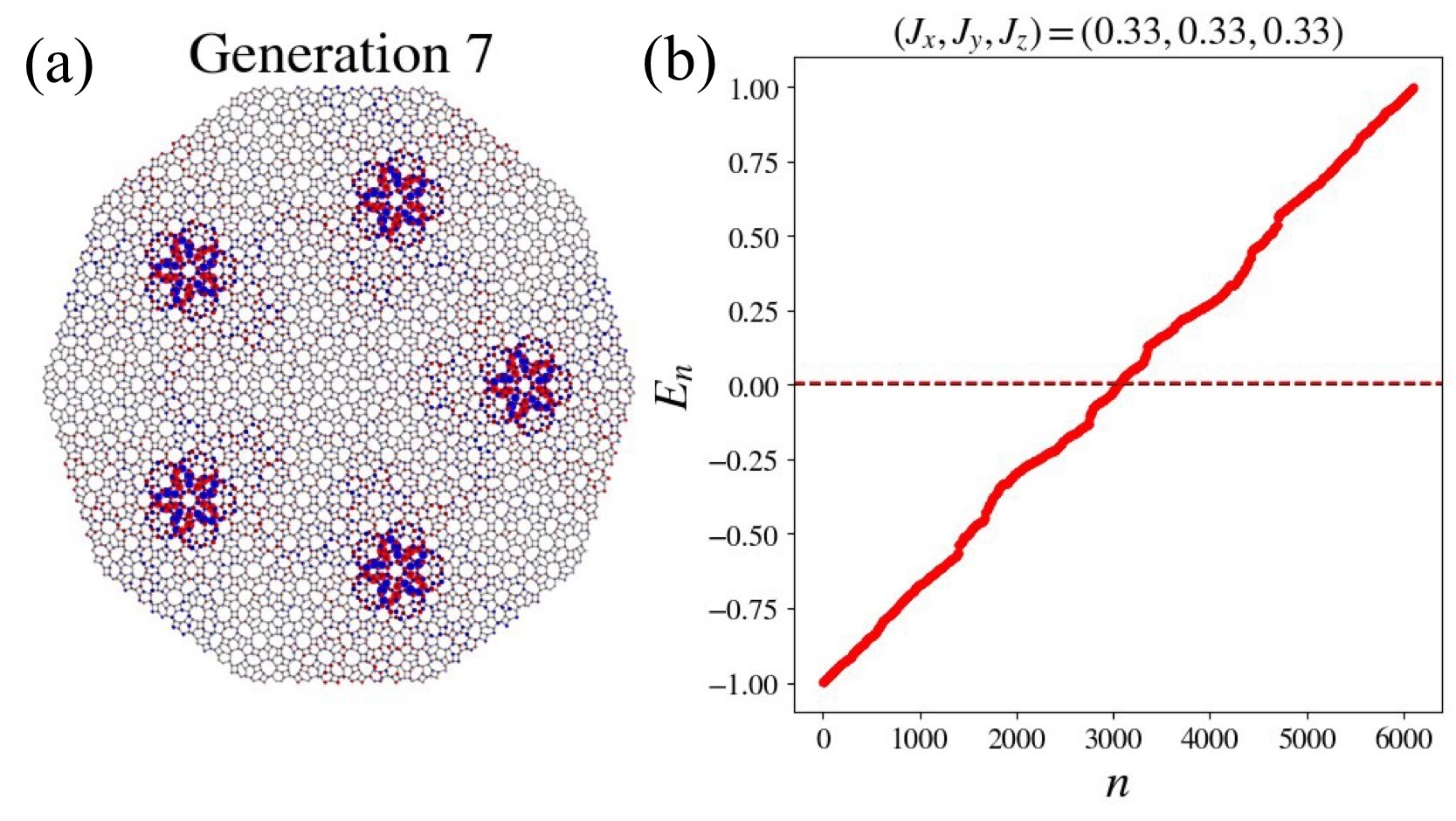}
\includegraphics[width=0.5\linewidth]{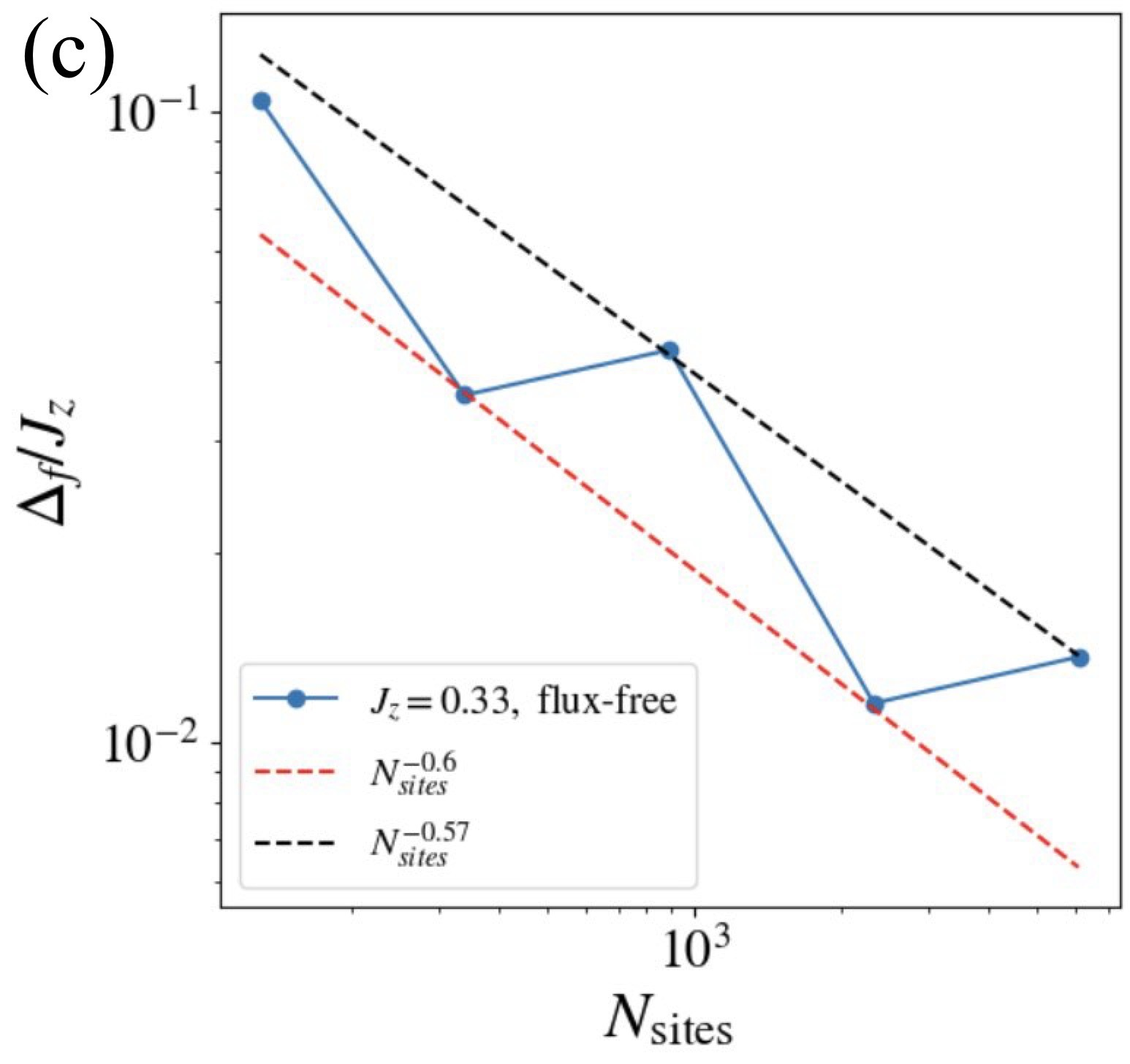}
\caption{Results for the higher energy flux-free sector. (a) The wave function distribution for the lowest-$E$ bulk state. (b) Fermionic spectrum. (c) Finite size scaling of the bulk gap. }
\label{fig:flux-free}
\end{figure}

We show in Fig.~\ref{fig:flux-free} results for the flux-free sector, which has higher energy than the canonical flux sector. In this case, the system at the isotropic point becomes gapless with $\Delta_f$ scaling as $\sim N_{\tn{sites}}^{-0.6}$. We note an even-odd effect manifest in the finite-size scaling of $\Delta_f$.

\section{LDOS as a function of position}
\label{appdx:ldos}
\begin{figure}[h]
\centering
\includegraphics[width=0.49\linewidth]{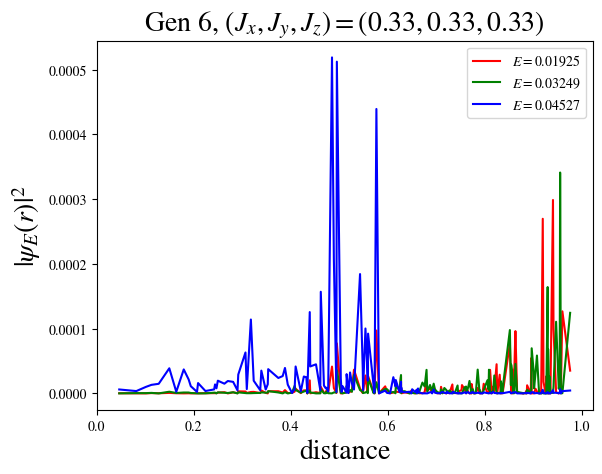}
\includegraphics[width=0.49\linewidth]{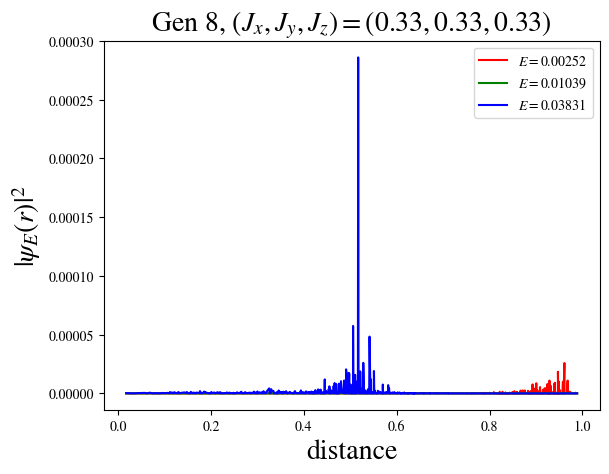}
\caption{Local DOS as a function of distance from the center. The distance is rescaled by the radius of the system.}
\label{fig:ldos}
\end{figure}

To identify true bulk excitations, we study the local density of states (LDOS) of low-energy excitations. For a given state with energy $E$, we can compute the amplitude of the state localized at site $i$, $\psi_E(i)$. Then we can obtain a histogram of $|\psi_E(r)|^2$ for all sites $i=1,\cdots,N_{sites}$ as a function of distance $r$ measured from the origin. In Fig.~\ref{fig:ldos}, we present the histogram for generations $\#$6, 8. The red and green curves are typical boundary states. The blue curves are the lowest-energy states for which the peaks are located deep in the bulk, and thus they can be identified as lowest-energy bulk states. Distance represents the atomic position rescaled by the radius of the system.

\section{Spectral functions of fermionic excitations in reciprocal space}

\begin{figure}[h]
\centering
\includegraphics[width=0.49\linewidth]{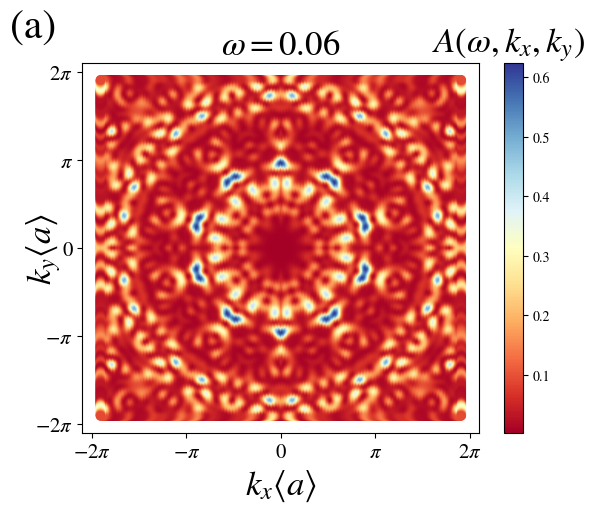}
\includegraphics[width=0.49\linewidth]{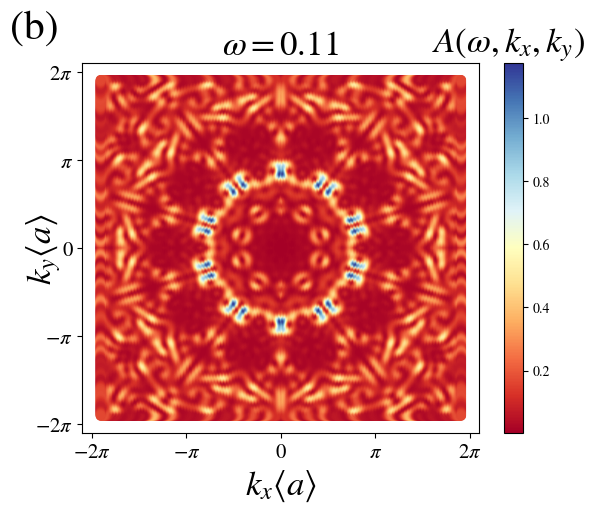}
\includegraphics[width=0.49\linewidth]{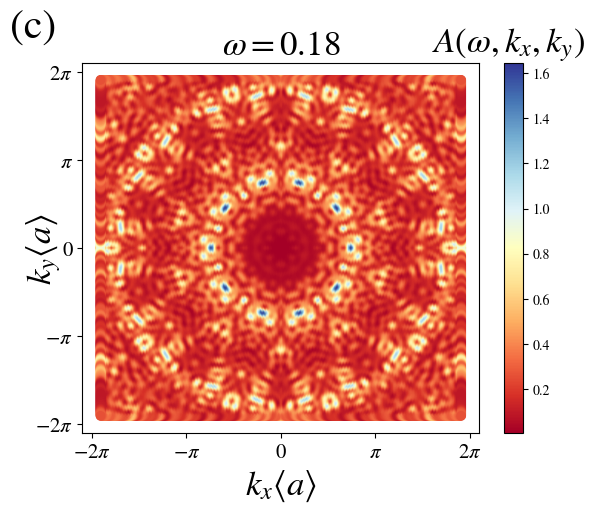}
\includegraphics[width=0.49\linewidth]{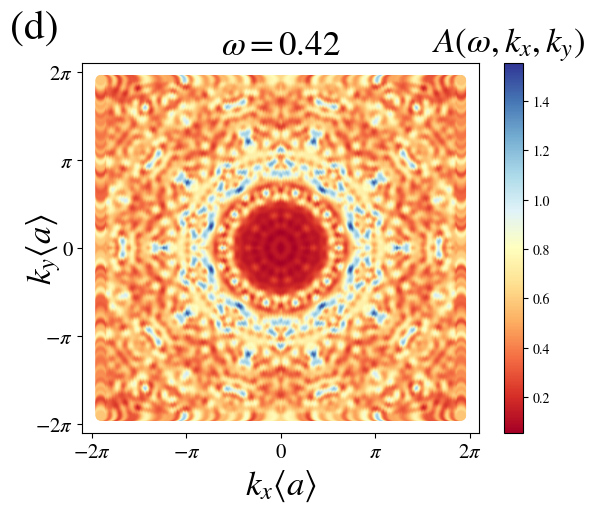}
\caption{Energy-resolved spectral function in  reciprocal space at four representative energy values (for generation $\#4$ at the isotropic point). $\langle a \rangle$ denotes an average atomic constant of the QC.}
\label{fig:QC_spectral_momentum}
\end{figure}

For a better understanding of the fermionic spectrum, we study the spectral function in reciprocal space. While the quasicrystal lacks a Brillouin zone associated with translational symmetry, we can still compute a spectral function defined as 
\beq 
A(\omega,\vec{k})&=&-\frac{1}{\pi}\tn{Im}G^{R}(\omega,\vec{k})\nonumber\\
&=&-\frac{1}{\pi}\tn{Im}\sum_{\vec{r}_1, \vec{r}_2} \psi_{\vec{k}}(\vec{r}_1)^*G^{R}(\omega,\vec{r}_1,\vec{r}_2)\psi_{\vec{k}}(\vec{r}_2),
\label{eq:spectral_fn}
\eeq
where $G^{R}(\omega,\vec{r}_1,\vec{r}_2)$ is the retarded Green's function and $\psi_{\vec{k}}(\vec{r})=\frac{1}{\sqrt{N}}e^{i\vec{k}\cdot\vec{r}}$ is the plane wave function. In Fig.~\ref{fig:QC_spectral_momentum}, we present energy-resolved spectral functions in  reciprocal space at four representative energy values across the spectrum at the isotropic point for generation $\#4$. The spectral function displays salient differences from the counterpart on the honeycomb model. First, the spectral functions exhibit a clear 5-fold rotational symmetry tied to the underlying symmetry of the quasicrystal. Moreover, there are 10 peaks of the spectral function at finite energy $\omega>\Delta_f$, around which the Dirac-like features are absent.

\normalfont
\bibliographystyle{apsrev4-1_custom}
\bibliography{References}

\clearpage

\begin{widetext}

\newcommand{\supplementary}{%
  \par
  \setcounter{section}{0}%
  \renewcommand{\thesection}{\Roman{section}}
  \renewcommand{\theHsection}{supplementary.\thesection}
  \titleformat{\section}[block]{\normalfont\bfseries\centering}{\thesection.}{1em}{}%
}
\makeatother
\supplementary
\setcounter{page}{1} 
\setcounter{equation}{0} 
\renewcommand{\figurename}{Supplemental Figure}
\renewcommand{\theequation}{\thesection.\arabic{equation}}

\begin{center}
    {\bf Supplementary material for ``Quasicrystalline Spin Liquids"}\\
    Sunghoon Kim, Mohammad Saad, Dan Mao, Adhip Agarwala, Debanjan Chowdhury
\end{center}

\section{Quasicrystalline graphs with an alternative choice of colors}

\renewcommand{\thefigure}{S1}
\begin{figure}[h]
\includegraphics[width=0.33\linewidth]{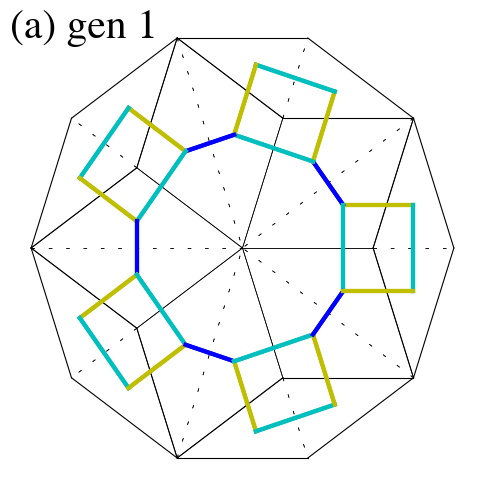}
\includegraphics[width=0.33\linewidth]{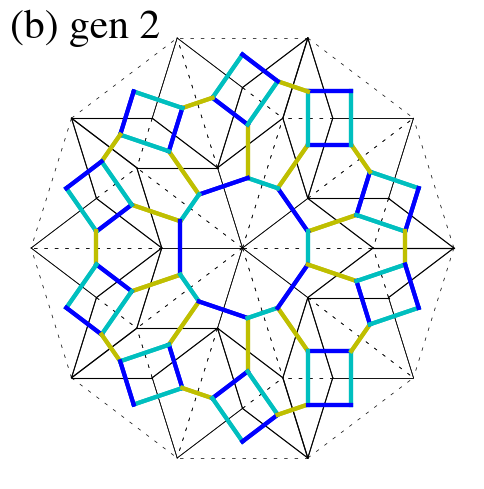}
\includegraphics[width=0.33\linewidth]{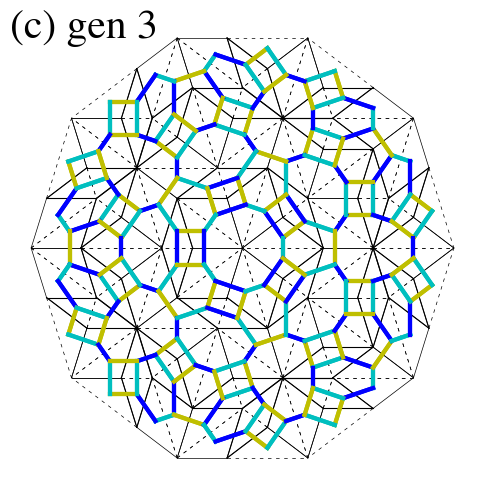}
\caption{Construction of our quasicrystals using an alternative set of colors for each bond}
\label{fig:new_coloring}
\end{figure}
We present our tri-coordinated quasicrystals for the first three generations with an alternative set of colors in Fig.~\ref{fig:new_coloring}.

\end{widetext}
\end{document}